\documentclass[
reprint,
superscriptaddress,
amsmath,
amssymb,
aps,
prl
]{revtex4-2}

\usepackage[T1]{fontenc}
\usepackage[utf8]{inputenc}
\usepackage{graphicx}
\usepackage{bm}
\usepackage[normalem]{ulem}
\usepackage{xcolor}
\usepackage[
colorlinks,
citecolor=blue,
linkcolor=blue,
urlcolor=blue
]{hyperref}

\begin{document}

\title{Simplified criteria for identifying and designing altermagnetic crystals in real space}

\author{Ying Chen}
\affiliation{Eastern Institute of Technology, Ningbo 315200, China}
\affiliation{School of Physical Sciences, University of Science and Technology of China, Hefei 230026, China}

\author{Qiushi Huang}
\affiliation{Eastern Institute of Technology, Ningbo 315200, China}

\author{Yu Wu}
\affiliation{Advanced Thermal Management Technology and Functional Materials Laboratory, Ministry of Education Key Laboratory of NSLSCS, School of Energy and Mechanical Engineering, Nanjing Normal University, Nanjing 210023, China}

\author{Xiaolan Yan}
\affiliation{Eastern Institute of Technology, Ningbo 315200, China}

\author{Su-Huai Wei}
\email{suhuaiwei@eitech.edu.cn}
\affiliation{Eastern Institute of Technology, Ningbo 315200, China}

\begin{abstract}
Altermagnetism is a compensated magnetic phase characterized by zero net magnetization and exchange-driven spin splitting. However, identifying altermagnets among collinear antiferromagnets usually requires full magnetic-space-group or spin-group analysis, which is not always intuitive. Here we formulate a simple real-space criterion based on how the crystallographic operations of the host nonmagnetic structure permute the two opposite-spin sublattices. We show that altermagnets usually exist on collinear compensated antiferromagnets whose magnetic primitive cell coincides with the host nonmagnetic crystallographic primitive cell. In this case, altermagnetic spin splitting is generally allowed unless an inversion-type operation exists that exchanges the two opposite-spin sublattices. Using chemically ordered Mn$_2$SSe prototypes derived from zinc-blende or rocksalt parent structures, we demonstrate that these criteria can be easily used to construct the three symmetry classes by controlling chemical ordering and magnetic-sublattice permutation. Similar rules can also be applied to low-dimensional crystals or quasicrystals. Our work reduces the identification of altermagnets to a transparent real-space symmetry test and provides a practical route for designing altermagnetic crystals.
\end{abstract}

\maketitle

Altermagnets (AM) have recently emerged as a distinct class of collinear magnetically ordered materials, in which fully compensated antiparallel spin sublattices produce zero net magnetization while exchange-driven spin splitting persists even in the absence of spin-orbit coupling (SOC)\cite{Smejkal2022,Smejkal2022a,Fender2025,Belashchenko2025,Reimers2024,Hayami2019}. This unusual combination of zero global net magnetization and spin-polarized band structure has opened a route toward spin functionalities unavailable in conventional ferromagnets (FM) or spin-degenerate antiferromagnets (AFM)\cite{Zhang2025a,Bai2024,Lee2024,Smejkal2020,Wang2025,Chen2025}. Yet identifying altermagnets remains challenging and less intuitive, because it generally invokes knowledge of sophisticated magnetic-space-group or spin-group analysis, often together with co-representation theory for antiunitary symmetries\cite{Zeng2024a,Gallego2012,Bradley2009}. The resulting complexity has hindered systematic searches or design of new altermagnetic materials and motivates us to provide a more direct criterion based on real-space group operations \cite{Gao2025,Guo2023}.

Here we formulate such a criterion for collinear fully compensated antiferromagnets by examining how the crystallographic operations of the host nonmagnetic structure (i.e., when the spin character is removed) act on the two opposite-spin sublattice sites of the antiferromagnets or altermagnets. In the absence of SOC, spin is a good quantum number, i.e., the band energies can be labeled as $\varepsilon_n(\mathbf{k},\uparrow)$ and $\varepsilon_n(\mathbf{k},\downarrow)$, where $\mathbf{k}$ is the momentum. Because the system is globally antiferromagnetic, the system must have an equal number of crystallographically equivalent spin-up and spin-down sites, forming two spin-sublattices. If the spin-cell is a supercell of the host nonmagnetic cell and the spin-up and spin-down sublattices are connected simply by the host translational vector $T$, then $\varepsilon_n(\mathbf{k},\uparrow)=\varepsilon_n(\mathbf{k},\downarrow)$, because $R_T\mathbf{k}=\mathbf{k}$, i.e., the bands are spin-degenerate at all $\mathbf{k}$ points\cite{Cheong2025,Yuan2020a} (see more detailed discussion below). Therefore, in this study, we only focus on systems whose magnetic primitive cell coincides with the nonmagnetic crystallographic primitive cell. For brevity, we refer to them as equal magnetic and nonmagnetic unit cell (EMNUC) systems below.

Within this class, generic spin splitting is determined by whether an inversion-type symmetry exists and, if so, how it acts on the two opposite-spin sublattices. Same-$\mathbf{k}$ opposite-spin degeneracy throughout the Brillouin zone is enforced only when an inversion-type symmetry globally exchanges the two sublattices; otherwise, exchange-driven spin splitting is symmetry-allowed at generic momenta. This criterion provides both a transparent real-space diagnostic and a constructive design principle. Chemical ordering of mixed anions or mixed cations can enlarge the primitive cell and remove spin-exchanging primitive translations. When inversion is retained, the magnetic arrangement determines how it acts on the spin sublattices. Guided by this criterion, we design chemically ordered Mn$_2$SSe prototypes that realize all three global symmetry classes within a single chemical composition and establish how little-group symmetries govern same-$\mathbf{k}$ opposite-spin degeneracy in momentum space.

\textit{Real-space criterion for spin degeneracy---}The criterion is most transparently formulated by tracking how crystallographic operations of the host nonmagnetic structure act on the two opposite-spin sublattices. As stated above, for a collinear antiferromagnet in the absence of SOC, spin remains a good quantum number, allowing the band energies to be labeled as $\varepsilon_n(\mathbf{k},\uparrow)$ and $\varepsilon_n(\mathbf{k},\downarrow)$.

Let $G$ be the space group of the nonmagnetic crystallographic structure and let $\mathcal{L}_{\uparrow}$ and $\mathcal{L}_{\downarrow}$ denote the two sets of magnetic sites carrying opposite collinear moments. Once the antiferromagnetic spin pattern is specified, we first distinguish two spin-compatible classes according to their action on the magnetic sublattices:
\begin{equation}
G^{+}:\quad
g\mathcal{L}_{\uparrow}=\mathcal{L}_{\uparrow},\quad
g\mathcal{L}_{\downarrow}=\mathcal{L}_{\downarrow},
\end{equation}
\begin{equation}
G^{-}:\quad
g\mathcal{L}_{\uparrow}=\mathcal{L}_{\downarrow},\quad
g\mathcal{L}_{\downarrow}=\mathcal{L}_{\uparrow}.
\end{equation}
Here $G^{+}$ contains operations that preserve each spin sublattice, whereas $G^{-}$ contains operations that globally exchange the two opposite-spin sublattices. For the antiferromagnetic configurations considered here, the two opposite-spin sublattices are related by at least one crystallographic operation, and hence $G^{-}\neq 0$.

The sets $G^{+}$ and $G^{-}$ can be identified directly from the real-space spin pattern. We first treat the spin-up and spin-down magnetic atoms as two distinct atomic species, while leaving all nonmagnetic atoms unchanged. The crystallographic operations retained by the relabeled structure form $G^{+}$. The operations of the full crystallographic group $G$ are then examined to identify an element $g_{0}^{-}$ that maps every spin-up magnetic atom onto a spin-down atom and vice versa, while preserving the chemical identities. The union $G^{+}\cup G^{-}$ is closed under composition, and $G^{+}$ is an index-two subgroup of this union. Therefore, once one element $g_{0}^{-}\in G^{-}$ is identified, the entire set $G^{-}$ is obtained as the left or right coset of $G^{+}$, $G^{-}=g_{0}^{-}G^{+}=G^{+}g_{0}^{-}$, implying a one-to-one correspondence between $G^{+}$ and $G^{-}$.

In general, not every operation in $G$ belongs to either $G^{+}$ or $G^{-}$. Operations that preserve the spin labels of some magnetic atoms but exchange those of others form the residual set $G^{\mathrm{mix}}=G\setminus\left(G^{+}\cup G^{-}\right)$. Such operations transform the selected spin pattern into a distinct magnetic configuration, rather than into the original configuration or its globally spin-reversed partner. They are therefore not symmetries of the selected magnetic configuration and impose no Neumann constraints on its band structure. In the following, only $G^{+}$ and $G^{-}$ are relevant to the band-symmetry analysis.

Following Neumann's principle, an operation $g\in G^{+}$ imposes an equivalence within the same spin channel,
\begin{equation}
\varepsilon_n(\mathbf{k},\sigma)=\varepsilon_n(R_g\mathbf{k},\sigma),\qquad \sigma=\uparrow,\downarrow .
\end{equation}
For $g\in G^{-}$, the operation maps the spin-up sublattice onto the spin-down sublattice and imposes
\begin{equation}
\varepsilon_n(\mathbf{k},\uparrow)=\varepsilon_n(R_g\mathbf{k},\downarrow).
\end{equation}
This is the central relation of the present real-space criterion. In general, a crystallographic operation relates opposite-spin states at different momenta. Same-$\mathbf{k}$ spin degeneracy is enforced only when there exists a sublattice-exchanging operation $g\in G^{-}$ whose point operation maps $\mathbf{k}$ back to the same crystal momentum,
\begin{equation}
R_g\mathbf{k}=\mathbf{k}+\mathbf{K},
\end{equation}
where $\mathbf{K}$ is a reciprocal lattice vector.

A primitive translation $T$ is a special case because its point part is the identity and therefore leaves $\mathbf{k}$ unchanged. If $T$ maps one spin sublattice onto the other, then $\varepsilon_n(\mathbf{k},\uparrow)=\varepsilon_n(\mathbf{k},\downarrow)$ throughout the Brillouin zone. Such systems are ordinary spin-degenerate antiferromagnets. We therefore focus on EMNUC systems, where the magnetic primitive cell coincides with the nonmagnetic crystallographic primitive cell and no primitive translation of the nonmagnetic structure exchanges the two opposite-spin sublattices.

For EMNUC systems, inversion-type operations provide the remaining global diagnostic. An inversion-type operation $\widetilde{I}$ denotes either pure inversion $I$ or a translated inversion $\{I|\tau\}$. Its point part transforms $\mathbf{k}$ into $-\mathbf{k}$. In two-dimensional crystals, $C_{2z}$ effectively plays the role of inversion in momentum space, mapping $\mathbf{k}$ to $-\mathbf{k}$. If $\widetilde{I}$ exchanges the two opposite-spin sublattices,
\begin{equation}
\widetilde{I}\in G^{-},
\end{equation}
then the general $G^{-}$ relation gives
\begin{equation}
\varepsilon_n(\mathbf{k},\uparrow)=\varepsilon_n(-\mathbf{k},\downarrow).
\end{equation}
However, in the absence of SOC, each spin channel is described by a scalar-relativistic Hamiltonian with a real periodic potential, so the band energy in a fixed spin channel satisfies
\begin{equation}
\varepsilon_n(\mathbf{k},\sigma)=\varepsilon_n(-\mathbf{k},\sigma),\qquad \sigma=\uparrow,\downarrow .
\end{equation}
Combining these two equalities gives
\begin{equation}
\varepsilon_n(\mathbf{k},\uparrow)=\varepsilon_n(\mathbf{k},\downarrow)
\end{equation}
for all $\mathbf{k}$. Therefore, when an inversion-type operation exchanges the two opposite-spin sublattices, same-$\mathbf{k}$ spin degeneracy is enforced throughout the Brillouin zone. This case corresponds to an ordinary spin-degenerate antiferromagnet rather than an altermagnet.

This observation leads to a real-space classification of EMNUC antiferromagnetic configurations based on whether an inversion-type operation belongs to $G^{+}\cup G^{-}$ and, when it does, whether it preserves ($G^{+}$) or exchanges ($G^{-}$) the two opposite-spin sublattices.

\textit{Real-space classification---}According to the above criteria, EMNUC antiferromagnetic configurations can be divided into three cases.

\emph{Case I: no spin-compatible inversion-type operation---}The first case is that no inversion-type operation belongs to either $G^{+}$ or $G^{-}$. This includes both noncentrosymmetric structures, whose crystallographic group contains no inversion-type operation, and centrosymmetric structures in which an inversion-type operation maps the selected magnetic configuration onto neither itself nor its globally spin-reversed counterpart. In either situation, no inversion-type operation can enforce same-$\mathbf{k}$ opposite-spin degeneracy throughout the Brillouin zone. Although other sublattice-exchanging operations $g^{-}\in G^{-}$ exist, Eq.~(4) shows that they generally map a state at $\mathbf{k}$ to an opposite-spin state at a symmetry-transformed momentum $R_{g^{-}}\mathbf{k}$, which is not equivalent to $\mathbf{k}$ at a generic momentum. Therefore, exchange-driven spin splitting is symmetry-allowed at generic momenta, while same-$\mathbf{k}$ opposite-spin degeneracy can still be restored at special momenta where a sublattice-exchanging operation $g^{-}\in G^{-}$ satisfies $R_{g^{-}}\mathbf{k}=\mathbf{k}+\mathbf{K}$. Such momenta typically form high-symmetry points, lines, or planes, as discussed below in terms of the little-group constraint.

\emph{Case II: sublattice-preserving inversion-type operation---}The second case is that an inversion-type operation belongs to $G^{+}$,
\begin{equation}
\widetilde{I}\in G^{+}.
\end{equation}
Thus, $\widetilde{I}$ maps the selected magnetic configuration onto itself and preserves each of the two opposite-spin sublattices. According to Eq.~(3), it imposes only a same-spin $\mathbf{k}\leftrightarrow-\mathbf{k}$ equivalence, which is already contained in the fixed-spin relation of Eq.~(8). Because $\widetilde{I}$ does not exchange the two spin sublattices, it does not impose the opposite-spin relation in Eq.~(7) and therefore cannot enforce same-$\mathbf{k}$ opposite-spin degeneracy. Exchange-driven spin splitting is therefore symmetry-allowed at generic momenta. As in Case I, spin degeneracy may still be restored at special momenta if a sublattice-exchanging operation $g^{-}\in G^{-}$ satisfies the little-group condition.

\emph{Case III: sublattice-exchanging inversion-type operation---}The third case is that an inversion-type operation $\widetilde{I}$ belongs to $G^{-}$ and globally exchanges the two opposite-spin sublattices. As shown in Eqs.~(7)--(9), this enforces same-$\mathbf{k}$ opposite-spin degeneracy throughout the Brillouin zone. Consequently, exchange-driven spin splitting is forbidden in the absence of SOC, corresponding to an ordinary spin-degenerate antiferromagnet.

\begin{figure}[t]
\centering
\includegraphics[width=\linewidth]{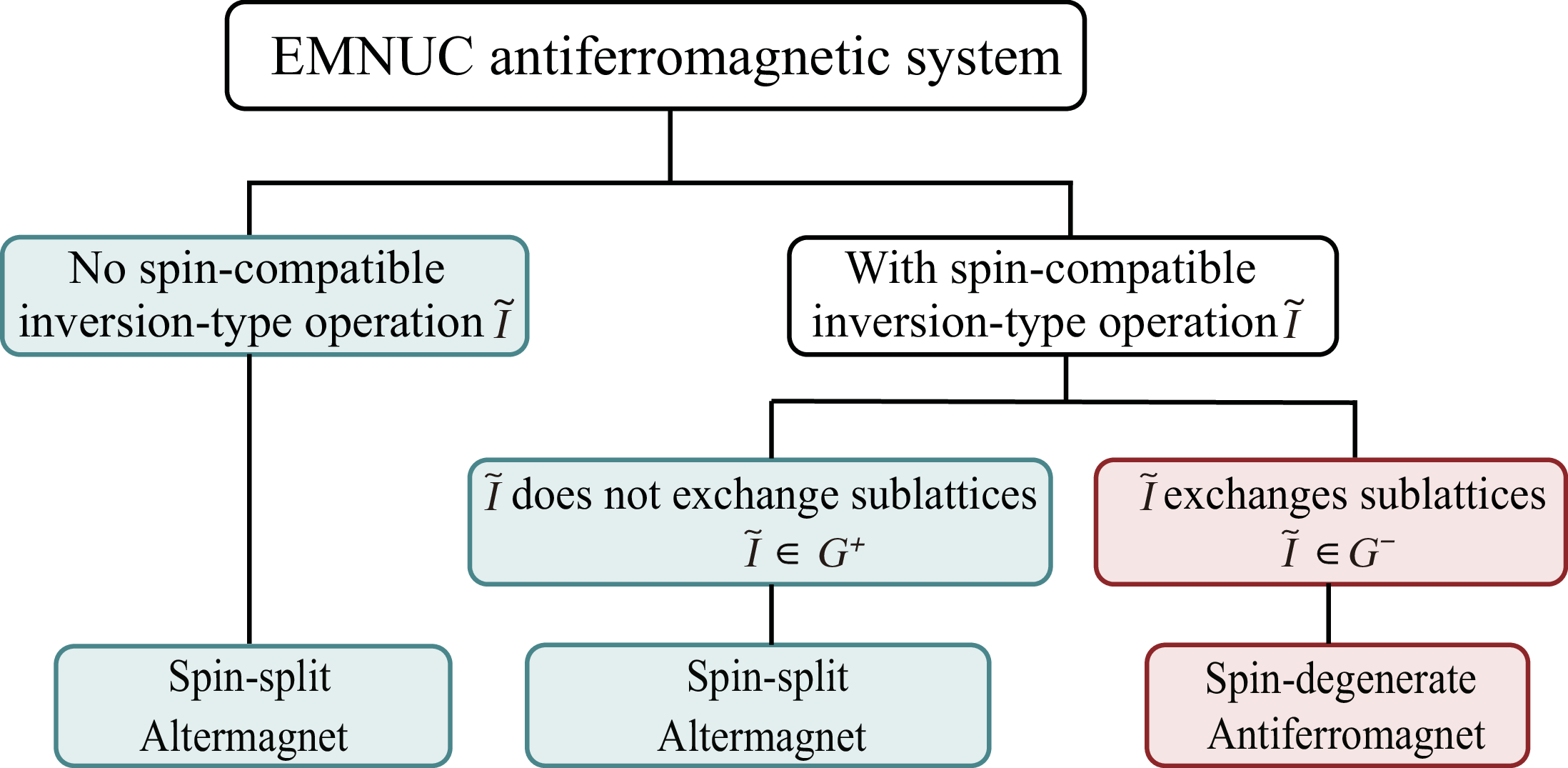}
\caption{Real-space classification of EMNUC antiferromagnetic configurations. Here an inversion-type operation $\widetilde I$ denotes either pure inversion $I$ or a translated inversion $\{I|\tau\}$ with $\tau\neq 0$. The left branch includes both structures without an inversion-type operation and structures in which an inversion-type operation belongs to $G^{\mathrm{mix}}$.}
\label{fig:type}
\end{figure}

The three cases are summarized in Fig.~1. Same-$\mathbf{k}$ spin degeneracy is globally enforced only in Case III, where an inversion-type operation belongs to $G^{-}$. In Cases I and II, exchange-driven spin splitting remains symmetry-allowed at generic momenta.

\textit{Little-group constraint---}The classification above determines whether same-$\mathbf{k}$ opposite-spin degeneracy is enforced throughout the Brillouin zone. In the altermagnetic Cases I and II, such degeneracy is not globally enforced, but it may still be restored at special momenta. Since operations in $G^{\mathrm{mix}}$ transform the selected spin pattern into a distinct magnetic configuration, they are excluded from the little group relevant to the magnetic state under consideration. We define
\begin{equation}
G_{\mathbf{k}}\equiv\left\{g\in G^{+}\cup G^{-}\,\middle|\,R_g\mathbf{k}=\mathbf{k}+\mathbf{K}\right\},
\end{equation}
where $\mathbf{K}$ is a reciprocal lattice vector.

Same-$\mathbf{k}$ opposite-spin degeneracy is enforced if
\begin{equation}
G_{\mathbf{k}}\cap G^{-}\neq 0.
\end{equation}
Indeed, for any $g'\in G_{\mathbf{k}}\cap G^{-}$, Eq.~(4) and the little-group condition give
\[
\varepsilon_n(\mathbf{k},\uparrow)=\varepsilon_n(R_{g'}\mathbf{k},\downarrow)=\varepsilon_n(\mathbf{k},\downarrow).
\]
Thus, although spin splitting is symmetry-allowed at generic momenta in Cases I and II, it is suppressed at special momenta whose little group contains a sublattice-exchanging operation.

If $G_{\mathbf{k}}\cap G^{-}=0$, all operations in $G_{\mathbf{k}}$ belong to $G^{+}$ and impose only same-spin equivalences; same-$\mathbf{k}$ spin splitting therefore remains symmetry-allowed. A sublattice-exchanging operation $g''\in G^{-}$ outside $G_{\mathbf{k}}$ may still relate the two spin channels through Eq.~(4). In that case, one obtains
\[
\varepsilon_n(\mathbf{k},\uparrow)=\varepsilon_n(R_{g''}\mathbf{k},\downarrow),
\]
where $R_{g''}\mathbf{k}$ is not equivalent to $\mathbf{k}$. The opposite-spin counterpart of a state at $\mathbf{k}$ therefore appears at a symmetry-transformed momentum rather than at the same $\mathbf{k}$. This momentum-separated opposite-spin partner relation is a characteristic feature of altermagnetic band structures.

\textit{Representative materials and constructive design---}We now apply the real-space criterion to two Mn$_2$SSe mixed anion structural prototypes derived from the zinc-blende and rocksalt Mn$X$ parent structures. The primitive cell of each parent contains only one Mn site and therefore cannot accommodate compensated magnetic sublattices within an EMNUC configuration\cite{Wei1986,MP}. Introducing an ordered periodic S/Se occupation pattern on the anion sublattice enlarges the crystallographic primitive cell and provides the multiple Mn sites required for the compensated AM and AFM types considered below.

In the zinc-blende-derived prototype, alternating S- and Se-containing anion layers along [001] yields a CuAu-I-type structure with a noncentrosymmetric primitive cell containing two Mn sites. In the rocksalt-derived prototype, a chalcopyrite-type S/Se occupation pattern yields a centrosymmetric primitive cell containing four Mn sites, which can accommodate distinct compensated AM and AFM types. Following the conventional prototype nomenclature for isovalent semiconductor alloys, we refer to these structures as zinc-blende-derived CuAu-I-type and rocksalt-derived chalcopyrite-type Mn$_2$SSe, respectively\cite{MP,Wei1987CuAu,Wei1989}.

\begin{figure}[t]
\centering
\includegraphics[width=\linewidth]{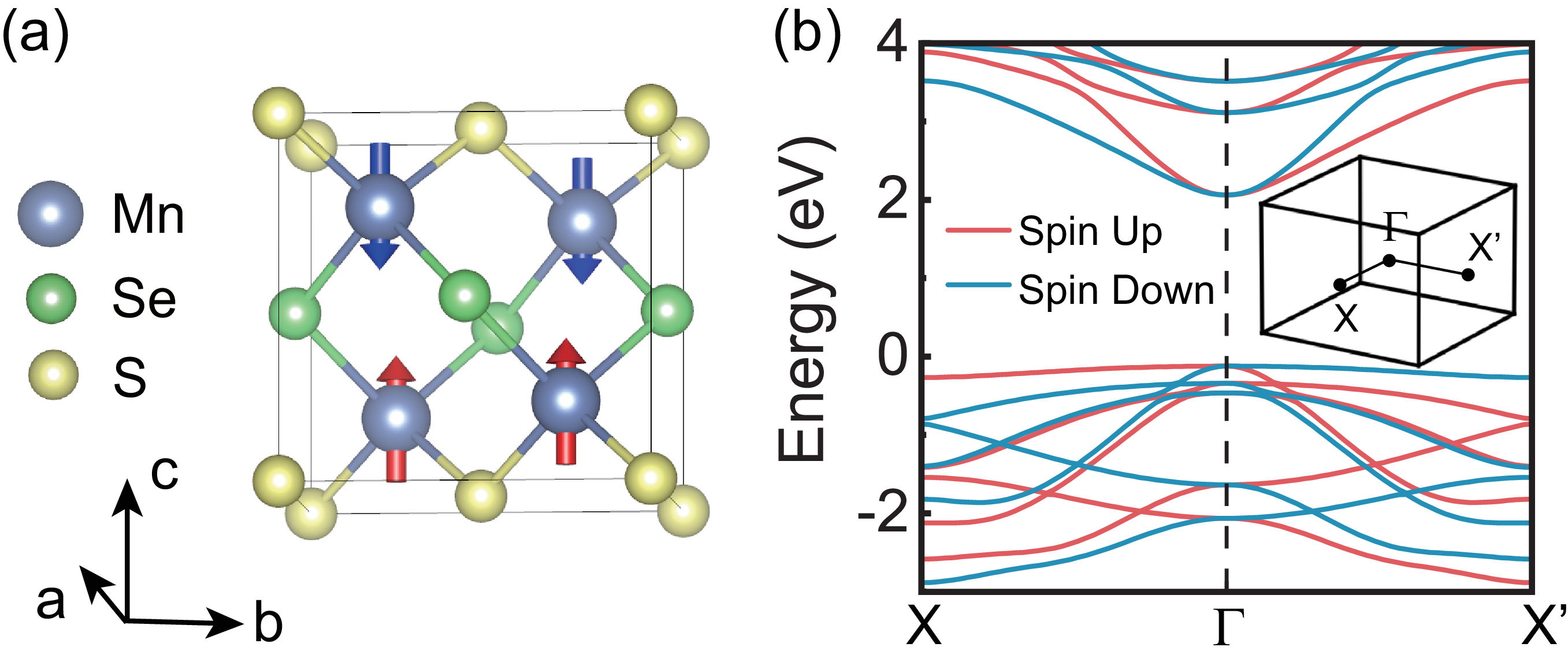}
\caption{CuAu-I-type Mn$_2$SSe. (a) Crystal and magnetic structure. For clarity, the structure is shown using an enlarged $\sqrt{2}\times\sqrt{2}\times1$ cell; the symmetry analysis and band calculation use the four-atom primitive cell shown in the Supplemental Material. Blue and red arrows denote the collinear Mn magnetic moments. (b) Spin-resolved band structure calculated without SOC. The representative high-symmetry points are $X=(0,1/2,0)$ and $X'=(1/2,0,0)$ in the corresponding primitive reciprocal basis~\cite{Hinuma2017}.}
\label{fig:cuau}
\end{figure}

\emph{Case I Zinc-blende-derived CuAu-I-type Mn$_2$SSe---}The crystal structure of CuAu-I-type Mn$_2$SSe is shown in Fig.~\ref{fig:cuau}(a). It belongs to the noncentrosymmetric space group P$\bar{4}$m2 with point group $D_{2d}$\cite{qiushicode,Wang2021VASPKIT}. Because no inversion-type operation $\widetilde I=I$ or $\widetilde I=\{I|\tau\}$ exists, no global inversion-based operation can enforce same-$\mathbf{k}$ opposite-spin degeneracy. Its crystallographic operations are divided into $G^{+}$ and $G^{-}$ in Table~\ref{tab:operation_classes}, while their explicit Seitz forms and coordinate transformations are given in the Supplemental Material.

\begin{table}[t]
\caption{Classification of the spatial symmetry operations into the spin-preserving subgroup $G^{+}$ and the spin-exchanging coset $G^{-}$. The point-operation symbols are expressed in the conventional tetragonal axes, with $x\parallel a$, $y\parallel b$, and $z\parallel c$; the fractional translations are omitted and listed in the Supplemental Material. Here, $C_2'=\{C_{2x},C_{2y}\}$, $C_2''=\{C_{2,[110]},C_{2,[1\bar{1}0]}\}$, $2\sigma_v=\{\sigma_{xz},\sigma_{yz}\}$, $2\sigma_d=\{\sigma_{x=y},\sigma_{x=-y}\}$, and $S_{4z}^{\pm}=\sigma_h C_{4z}^{\pm}$.}
\label{tab:operation_classes}
\centering
\renewcommand{\arraystretch}{1.25}
\setlength{\tabcolsep}{0pt}
\begin{tabular*}{\columnwidth}{@{\extracolsep{\fill}}lll@{}}
\hline\hline
\shortstack[l]{Magnetic\\type} & \shortstack[l]{$G^{+}$\\operations} & \shortstack[l]{$G^{-}$\\operations} \\
\hline
\shortstack[l]{CuAu-I-type\\AM} & \shortstack[l]{$E$, $C_{2z}$,\\$2\sigma_v$} & \shortstack[l]{$S_{4z}^{\pm}$,\\$2C_2''$} \\[1.0ex]
\shortstack[l]{Chalcopyrite-\\type AM} & \shortstack[l]{$E$, $C_{2z}$, $I$, $\sigma_h$,\\$2\sigma_v$, $2C_2'$} & \shortstack[l]{$C_{4z}^{\pm}$, $S_{4z}^{\pm}$,\\$2C_2''$, $2\sigma_d$} \\[1.3ex]
\shortstack[l]{Chalcopyrite-\\type AFM} & \shortstack[l]{$E$, $C_{2z}$, $C_{4z}^{\pm}$,\\$2\sigma_v$, $2\sigma_d$} & \shortstack[l]{$I$, $\sigma_h$, $S_{4z}^{\pm}$,\\$2C_2'$, $2C_2''$} \\
\hline\hline
\end{tabular*}
\end{table}

Along the representative $\Gamma$--$X$ direction, $\mathbf{k}=(0,u,0)$, a generic point with $0<u<\frac12$ has little group $C_s$, containing only
\[
E,\qquad \sigma_v(\sigma_{yz}),
\]
both of which belong to $G^{+}$. At the endpoint $X=(0,\frac12,0)$, the little group is $C_{2v}$,
\[
G_X=\{E,C_{2z},2\sigma_v\},
\]
which again contains only operations from $G^{+}$. No operation in $G^{-}$ belongs to the little group at a generic point on the line or at the endpoint $X$. Therefore, opposite-spin degeneracy is not symmetry enforced at these momenta. At $\Gamma$, by contrast, the little group contains operations from $G^{-}$ and restores the opposite-spin degeneracy. In addition, the operation $C_{2,[110]}\in G^{-}$ maps $(0,u,0)$ onto $(u,0,0)$ and relates the opposite-spin branches on the symmetry-related $\Gamma$--$X$ and $\Gamma$--$X'$ lines. The calculated spin-resolved bands in Fig.~\ref{fig:cuau}(b) therefore realize the Case I altermagnetic splitting.

\begin{figure}[t]
\centering
\includegraphics[width=\linewidth]{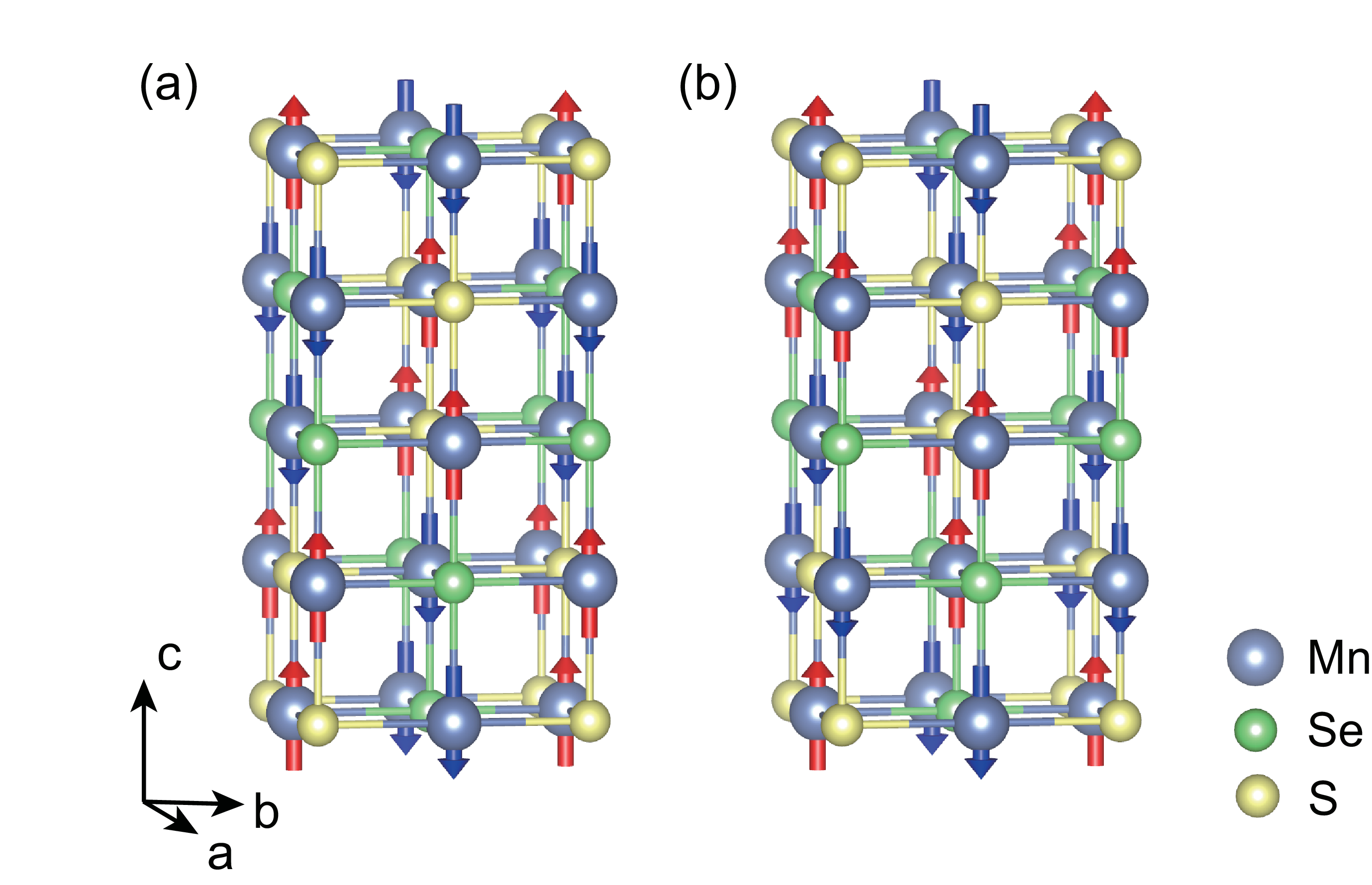}
\caption{Crystal and magnetic structures of chalcopyrite-type Mn$_2$SSe. (a) AM configuration. (b) AFM configuration.}
\label{fig:chalcopyrite_structure}
\end{figure}

\emph{Case II AM and Case III AFM configurations in centrosymmetric chalcopyrite-type Mn$_2$SSe---}The conventional cell of rocksalt-derived chalcopyrite-type Mn$_2$SSe is shown in Fig.~\ref{fig:chalcopyrite_structure}. It belongs to the centrosymmetric body-centered tetragonal space group $I4_1/amd$ with point group $D_{4h}$\cite{qiushicode,Wang2021VASPKIT}. The electronic-structure calculations use an eight-atom primitive cell with the lattice-vector basis defined in Ref.~\cite{Jaffe1983}, and the corresponding structural information is provided in the Supplemental Material. The AM and AFM spin patterns are shown in Figs.~\ref{fig:chalcopyrite_structure}(a) and \ref{fig:chalcopyrite_structure}(b), respectively. In the AM configuration, the inversion-type operation preserves each spin sublattice, whereas in the AFM configuration it exchanges the two spin sublattices:
\[
\begin{aligned}
\text{AM:}\quad&
\widetilde I\mathcal{L}_{\uparrow}=\mathcal{L}_{\uparrow},\qquad
\widetilde I\mathcal{L}_{\downarrow}=\mathcal{L}_{\downarrow},\qquad
\widetilde I\in G^{+},\\
\text{AFM:}\quad&
\widetilde I\mathcal{L}_{\uparrow}=\mathcal{L}_{\downarrow},\qquad
\widetilde I\mathcal{L}_{\downarrow}=\mathcal{L}_{\uparrow},\qquad
\widetilde I\in G^{-}.
\end{aligned}
\]
Here $\widetilde I=\{I|(0,\frac12,\frac14)\}$ in the conventional fractional coordinates. For the AM configuration, we consider the representative $\Gamma$--$N$ path, where $N=(1/2,0,1/2)$ in the primitive reciprocal basis used in the calculation. At a generic momentum along this path, the little group is $C_s=\{E,\sigma_v(\sigma_{xz})\}$, which contains only operations from $G^{+}$. Therefore, $G_{\mathbf{k}}\cap G^{-}=0$, and same-$\mathbf{k}$ opposite-spin degeneracy is not enforced. At the boundary point $N$, the little group is enlarged to $C_{2h}=\{E,C_2'(C_{2y}),I,\sigma_v(\sigma_{xz})\}$, which also contains no $G^{-}$ operation for AM configuration. Hence, spin splitting remains symmetry-allowed at generic momenta along $\Gamma$--$N$ and the endpoint $N$. At $\Gamma$, by contrast, operations in $G^{-}$ enter the little group and restore opposite-spin degeneracy.

\begin{figure}[t]
\centering
\includegraphics[width=\linewidth]{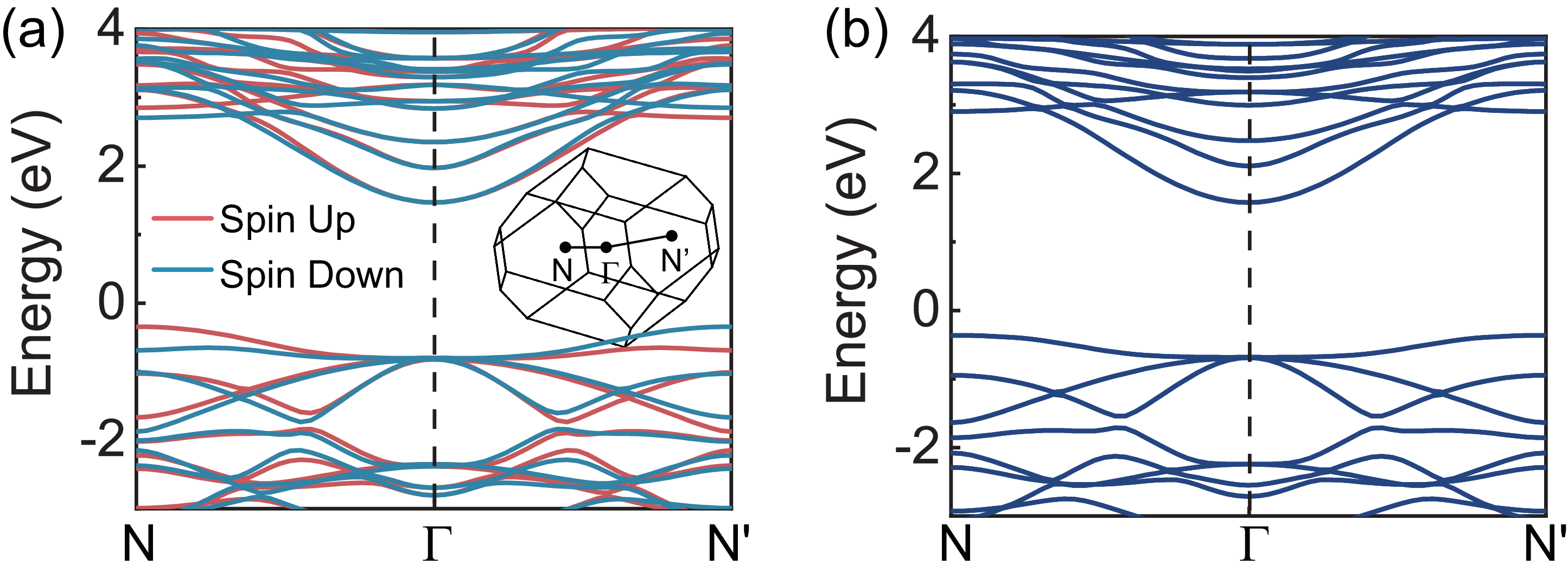}
\caption{Spin-resolved band structures of chalcopyrite-type Mn$_2$SSe calculated without SOC. (a) Case II AM configuration. (b) Case III AFM configuration. The Brillouin-zone boundary endpoints are $N=(1/2,0,1/2)$ and $N'=(0,1/2,1/2)$, in the primitive reciprocal basis used in the calculation.}
\label{fig:chalcopyrite_bands}
\end{figure}

The $G^{-}$ operation whose point part is $\sigma_d$ ($\sigma_{x=y}$) maps this path onto the symmetry-related $\Gamma$--$N'$ path, where $N'=(0,1/2,1/2)$, while exchanging the two spin sublattices. The two paths therefore exhibit opposite spin characters, as shown in Fig.~\ref{fig:chalcopyrite_bands}(a).

For the AFM configuration, the inversion-type operation exchanges the two spin sublattices, so that $\widetilde I\in G^{-}$. It therefore enforces same-$\mathbf{k}$ opposite-spin degeneracy throughout the Brillouin zone, consistent with the spin-degenerate bands in Fig.~\ref{fig:chalcopyrite_bands}(b).

Together with the noncentrosymmetric CuAu-I-type prototype, the two chalcopyrite-type spin configurations realize all three cases of the real-space criterion within the same Mn$_2$SSe composition. Chemical ordering creates the required magnetic sites and determines the available crystallographic operations. The spin pattern then determines whether an inversion-type operation, when present, belongs to $G^{+}$, $G^{-}$, or neither.

\textit{Conclusion---}We have established real-space symmetry criteria for altermagnets in EMNUC collinear compensated antiferromagnets following Neumann's principle, $\varepsilon_n(\mathbf{k},\sigma)=\varepsilon_n(R_g\mathbf{k},g\sigma)$. In this case, global same-$\mathbf{k}$ opposite-spin degeneracy is normally absent unless an inversion-type operation exchanges the two opposite-spin sublattices, i.e., the combined spin-inversion operation restores the equivalence between the two spin channels and forbids altermagnetic splitting in the absence of SOC. At high-symmetry momenta, the little group further determines whether the splitting survives locally. Using zinc-blende-derived CuAu-I-type and rocksalt-derived chalcopyrite-type Mn$_2$SSe prototypes as examples, we demonstrate the three symmetry classes within a single chemical composition: a noncentrosymmetric Case I altermagnet, a centrosymmetric Case II altermagnet, and a centrosymmetric Case III spin-degenerate antiferromagnet. Our analysis, therefore, provides not only a transparent real-space identification and classification rule, but also a practical design principle for engineering altermagnetic crystals. Such a criterion can be easily extended to other systems such as low-dimensional materials and quasicrystals.

\textit{Acknowledgments---}This work was supported by the National Key Research and Development Program of China (Grant No. 2024YFA1409800), the National Natural Science Foundation of China (Grant No. 12688201) and the China Postdoctoral Science Foundation (Grant No. 412519.01.00.011).

\end{document}